\documentclass[12pt]{article}
\usepackage{graphicx}
\usepackage[hidelinks]{hyperref}

\usepackage{amsmath,amsfonts,amsfonts,amssymb,amsthm}

\usepackage{geometry,ulem,setspace}

\usepackage{xurl}

\usepackage{caption}
\usepackage{subcaption}

\usepackage{pgfplots}
\usetikzlibrary{patterns}

\usepackage[round]{natbib}
\usepackage{outlines}
\usepackage{titling}

\usepackage{amsthm}

\usepackage{titlesec}
\titleformat*{\section}{\large\bfseries}

\usepackage{tcolorbox}
\usepackage{blindtext}
\newtcolorbox[auto counter]{mybox}[2][]{
    colback=black!5!white,
    colframe=black!40!white,
    colbacktitle=black!20!white,
    coltitle=black,
    center,
    title={Box~\thetcbcounter: #2},
    #1
}

\title{Undermining Mental Proof: How AI Can Make Cooperation Harder by Making Thinking Easier}

\author{Zachary Wojtowicz\thanks{Corresponding Author.} \thanks{Department of Economics and Harvard Business School, Harvard University, Cambridge, MA, United States; {\tt zwojtowicz@fas.harvard.edu}} and Simon DeDeo\thanks{Dept of Social and Decision Sciences, Carnegie Mellon University, Pittsburgh, PA \& the Santa Fe Institute, Santa Fe, NM; {\tt sdedeo@andrew.cmu.edu}}}
\date{\today}

\begin{document}

\maketitle

\begin{abstract}
Large language models and other highly capable AI systems ease the burdens of deciding what to say or do, but this very ease can undermine the effectiveness of our actions in social contexts. We explain this apparent tension by introducing the integrative theoretical concept of ``mental proof,'' which occurs when observable actions are used to certify unobservable mental facts. From hiring to dating, mental proofs enable people to credibly communicate values, intentions, states of knowledge, and other private features of their minds to one another in low-trust environments where honesty cannot be easily enforced. Drawing on results from economics, theoretical biology, and computer science, we describe the core theoretical mechanisms that enable people to effect mental proofs. An analysis of these mechanisms clarifies when and how artificial intelligence can make low-trust cooperation harder despite making thinking easier.
\end{abstract}

%

\section{Introduction}

The widespread availability of generative artificial intelligence means that anyone can now cheaply and convincingly simulate the output of human mental effort across an unprecedented variety of tasks. This promises numerous benefits across nearly every aspect of society, but it also has begun to disrupt an equally broad set of social practices, such as sincere apologies \citep{glikson2023ai}, college assessment \citep{fitria2023artificial,cardon2023challenges}, online dating \citep{wu2020online}, and wedding vows \citep{nyt}. 

In light of these developments, many have come to see the technology as a double-edged sword: artificial intelligence cuts the cost of thinking, but it also---and, as we will argue, \emph{for that very reason}---threatens vital elements of the existing social fabric, such as trust \citep{glikson2020human}, privacy \citep{jain2023contextual}, public safety \citep{leslie2019understanding}, and even democracy itself \citep{allen2024real,jungherr2023artificial}. Indeed, a majority of survey respondents now feel ``more concerned than excited'' about the increased use of artificial intelligence in their daily lives \citep{Tyson_2023}.


Despite the urgency of these concerns, the scientific community has struggled to articulate general principles that explain why lowering mental costs through artificial intelligence undermines such a wide variety of seeming unrelated social practices. We may ``know it when we see it'' in any particular case, but the lack of integrative scientific frameworks has made it difficult to formulate general solutions to existing problems or foresee future harms. 

In this paper, we highlight the key role that ``mental proof'' plays in facilitating cooperation in low-trust environments. Mental proofs are observable actions taken to certify unobservable facts about the minds who perform them. 
As we describe fully below, people use two distinct mechanisms to substantiate mental proofs:
signaling theory (an idea primarily studied in economics and biology) and proof of knowledge protocols (studied in computer science). To function, both mechanisms rely on implicit assumptions about the cost structure of organic mental activity---assumptions that are rapidly being disrupted by the proliferation of artificial intelligence in daily life.

An appreciation for the structure of mental proofs therefore helps elucidate the underlying logic of artificial intelligence's various social consequences. In highlighting the importance of mental proof and its relationship to thinking machines, our paper contributes to a wider, cross-disciplinary attempt to proactively understand and address the technology's various social consequences \citep[\emph{e.g.},][]{solaiman2023evaluating,weidinger2021ethical,mirsky2021creation}

We illustrate the practical importance of mental proof in two ``worked examples'' drawn from everyday social domains: sincere apology and subculture formation. In both cases, we discuss how low-cost simulations of intelligent behavior undermine the efficacy of mental proofs and the vital social benefits they provide.


As our discussion makes clear, mental proof is most valuable in situations where honesty cannot easily be enforced. This implies that the category of harms we discuss will disproportionately impact those who are not already embedded in high-trust networks and formal institutions, thereby reinforcing existing structural barriers to social mobility and economic development. Our analysis does, however, suggest effective strategies for mitigating these deleterious consequences, and we conclude with a sketch of the framework's implications for policy, technology, and everyday life.




\section{Mental Proof}

Communication can greatly enhance coordination. Our species' remarkable capacity for coordination relies, at its foundation, upon an ability to reliably externalize the nuances of our internal mental states---not just our beliefs, but also our intentions, values, preferences, skills, understandings, commitments, abilities, \emph{etc.}---in ways that others will not only understand, but trust. As has been pointed out by both behavioral scientists \citep[\emph{e.g.},][]{tomasello2005understanding} and philosophers of mind \citep[\emph{e.g},][]{gilbert1990walking,bratman1992shared}, people's ability to share and understand intentions, in particular, is essential to the formation of collaborative acts that range from going on a walk together to drafting a new constitution.




Sometimes this is easy: in many contexts, we can simply speak our minds and others will have good reason to believe us. Despite its many conveniences, however, ``cheap talk'' breaks down when people have incentives to lie \citep{farrell1987cheap}. In such contexts, mere assertions lose their credibility, and even those who try to tell the truth will be dismissed.

A variety of social institutions help enforce honesty and thereby preserve the benefits of its coordinating function, most notably reputation \citep{fehr2009economics}, norms \citep{elster1989social}, and formal punishment \citep{north1991institutions}. Unfortunately, however, these institutions are not always available---\emph{e.g.}, before reputation is established, when claims are difficult to verify, or in places where cultural and legal institutions are weak.

In these ``low-trust'' contexts, interacting parties must not only state, but endeavor to \emph{prove}, claims about their minds. One way people furnish such proof is by taking observable actions which (given certain assumptions about the capabilities and structure of human brains) provide strong grounds for a relevant claim about the mind. We refer to such behaviors as constituting \textbf{mental proof}. The validity of mental proofs are primarily underwritten by two separate mechanisms: \textbf{signaling theory} and \textbf{proof of knowledge protocols}. We review each, in turn.

\section{Mechanism One: Signaling Theory}

Signaling theory was introduced into economics by \cite{spence1973job} to explain how seemingly self-defeating behaviors (\emph{e.g.}, knowingly pursuing a degree in a field one is unlikely to use) can still benefit rational agents by signaling information about one's preferences or abilities (\emph{e.g}, that one is smart enough to graduate college) to others in a way that cannot be faked. Signaling was introduced to biology around the same time by \cite{zahavi1975mate} to explain an analogous class of animal traits and behaviors: phenotypes that seem to reduce fitness, such as peacocks growing elaborate tails or gazelles stotting.\footnotemark\ As the theory points out, these acts credibly communicate information to potential mates or predators precisely because of their self-handicapping effect; voluntarily wasting resources can credibly signal a wealth of resources to begin with.

\footnotetext{``Stotting'' refers to a behavior in which a quadruped acrobatically leaps into the air and contorts their body \citep{smith2003animal}. \cite{fitzgibbon1988stotting} document, for example, that: Thomson's gazelles are more likely to stott when wild dogs were nearby; that dogs were more likely to chase gazelles which stotted less frequently than their peers; and that gazelles which stotted more intensively had a better chance of escaping, conditional on being chased. This and other evidence has been taken to support the underlying hypothesis that such behaviors honestly signal a prey's capacity to outrun predators, which makes them less likely to give chase. This cooperative d\'{e}tente between adversaries saves both energy \citep{caro1994ungulate}.}
    
The central insight of signaling theory is that rational agents only take actions they expect to be beneficial on net. A behavior can constitute definitive proof that the actor expects its benefits (including revealing information to you, the potential observer) to outweigh its costs. This pinpoints why signaling can only be accomplished by behaviors that incur real net costs when faked. The apparent downside of signaling behaviors are precisely what establish their credibility: the cost structure ensures that deceptive types cannot send the signal with impunity. This self-policing logic is what enables signaling equilibria to extend credible communication to low-trust environments, where honesty can not be enforced (by, \emph{e.g}, reputation).

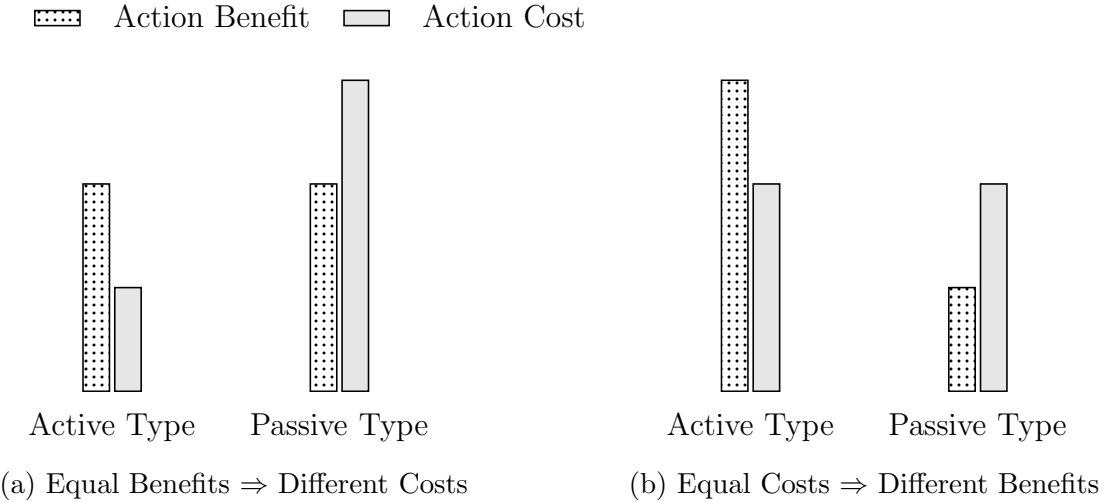
\begin{figure}[h!]
     \centering
     \begin{subfigure}[b]{0.5\textwidth}
         \centering
         \begin{tikzpicture}
            \begin{axis}[
                    width=1*\textwidth,
                    ybar,
                    enlarge x limits=.6,
                    ymin=0, ymax=6, 
                    axis line style={draw=none},
                    tick style={draw=none},
                    ytick=\empty,
                    symbolic x coords={Active Type, Passive Type},
                    xtick=data,
                    legend image code/.code={%
                        \draw[#1] (0cm,-0.1cm) rectangle (0.6cm,0.1cm);
                    },
                    legend style={
                        at={(0.1,0.65)},
                        anchor=west,
                        draw=none,
                        legend columns=-1,
                        column sep=2ex,
                    }
                ]
                \addplot[pattern = dots, semithick] coordinates {(Active Type,2) (Passive Type,2)}; 
                \addplot[draw=black, fill = black!10!white, semithick] coordinates {(Active Type,1) (Passive Type,3)}; 
                \legend {Action Benefit, Action Cost};
            \end{axis}
        \end{tikzpicture}
         \caption{Equal Benefits $\Rightarrow$ Different Costs}
         \label{fig:equal-benefit}
         \vspace{-5em}
     \end{subfigure}
     \begin{subfigure}[b]{0.5\textwidth}
         \centering
         \begin{tikzpicture}
            \begin{axis}[
                    width=1*\textwidth,
                    ybar,
                    enlarge x limits=.6,
                    ymin=0, ymax=6,
                    axis line style={draw=none},
                    tick style={draw=none},
                    ytick=\empty,
                    symbolic x coords={Active Type, Passive Type},
                    xtick=data,
                ]
                \addplot[pattern = dots, semithick] coordinates {(Active Type,3) (Passive Type,1)}; 
                \addplot[draw=black, fill = black!10!white, semithick] coordinates {(Active Type,2) (Passive Type,2)}; 
            \end{axis}
        \end{tikzpicture}
         \caption{Equal Costs $\Rightarrow$ Different Benefits}
         \label{fig:equal-cost}
     \end{subfigure}
        \caption{Two focal categories of signaling equilibria. In each, one ``active'' type of agent engages in a signaling behavior while the other ``passive'' type does not. To sustain a separating equilibrium, it must be that acting generates a net gain for the active type but a net loss for the passive type. In the first category of equilibrium (sub-figure a), both agent types receive the same benefit from engaging in a behavior. The fact that one type acts but the other does not implies that they incur different costs. In the second (sub-figure b), both types incur the same cost. Here, the fact that only one type acts implies that they expect different benefits.}
        \label{fig:costly-signaling}
\end{figure}

Figure~\ref{fig:costly-signaling} illustrates two focal cases of signaling equilibria. In the first case (subfigure a), two types of agents receive the same benefit from engaging in a signaling behavior, but incur different costs. From the perspective of an external observer, witnessing the behavior constitutes definitive proof that the agent is of the type with lower cost. Credibly communicating this distinction may benefit both parties, for example because having a low cost for the focal behavior (\emph{e.g.}, finding mathematics easy) correlates with other traits (\emph{e.g.}, abstract problem-solving ability) that are necessary for productive cooperation. 

In the second case (subfigure b), both agents incur the same cost, but receive different benefits. From the perspective of an external observer, witnessing such a behavior constitutes proof that the agent is of the high-benefit type. Credibly communicating this distinction can profit both parties, for example because it reveals that the person will want the cooperative relationship to continue further into the future.

In economic domains, costly signaling has been used to explain a wide variety of phenomena, including lavish, uninformative advertising efforts \citep{milgrom1986price} corporate social responsibility initiatives \citep{cheng2014corporate,flammer2021corporate}, and disclosures made by ventures seeking early-stage funding \citep{ahlers2015signaling}. Signaling is not only confined to ``western, education, industrial, rich, and democratic'' societies \citep{weird}, however; anthropologists have applied the concept to make sense of puzzling phenomena across the ethnographic record \citep{bliegebird2005signaling}. Biologists, for their part, have used costly signaling to explain cooperative alert calls \citep{bergstrom2001alarm}, ritual fighting \citep{zahavi1975mate}, singing \citep{mithen2006singing}, secondary sex characteristics such a bright coloring \citep{folstad1992parasites}, and a variety of other facts about the animal world.


Although, as the above examples make clear, any costly resource can underwrite the credibility of a signaling equilibrium, which resources happen to be relatively scarce or desirable will, of course, vary across times, places, and cultures.\footnotemark\ Mental resources are somewhat unique in this regard: in contrast to material goods, whose production depends on technology, capital, and other variable factors, the ``one person, one brain'' principle---\emph{i.e.}, fact that every individual is endowed a limited set of cognitive capacities \citep[such as working memory and cognitive control;][]{miller1956magical,shenhav2017toward} that cannot, on a physical level, be increased---ensures that the supply of human attention is essentially fixed \citep{loewenstein2023economics}.\footnotemark\

\footnotetext{\cite{bliegebird2005signaling} provide examples that range from particularly labor-intensive yams to turtle meat.}

\footnotetext{If anything, mental resources have become relatively more scarce over time in modern industrial economies as technology drives the cost of producing both material commodities \citep{nordhaus1996real} and information \citep{simon1971designing} toward zero.}



Many instances of social signaling rely on the fact that mental resources are scarce and therefore costly to expend. Mental activity entails both direct, metabolic costs \citep[\emph{e.g.}, firing neurons consumes oxygenated blood glucose and generates toxic metabolites;][]{attwell2001energy} and indirect, opportunity costs \citep[\emph{e.g.}, committing a limited-capacity resource such as cognitive control or working memory to one task precludes its simultaneous use for other, potentially valuable tasks;][]{shenhav2017toward}. On the one hand, this means that nearly every act of human intelligence incurs real costs; on the other, it means that all such acts have the potential to credibly signal one's preferences and abilities to others. 

The widespread commercial availability of highly capable artificial intelligence, however, has made it so that anyone can now quickly and cheaply simulate behaviors that previously required the concerted application of these mental resources. A simple example, discussed in detail in Section~\ref{sec:apology}, is the act of writing a sincere apology---a mentally effortful activity that can now be automated with a simple prompt.
By driving the cost of these mental activities to zero, artificial intelligence has the potential to undermine the very possibility of using signaling as a vehicle for establishing mental proofs across an increasingly wide variety of domains. 

Signaling theory predicts that the extent of these effects will depend upon the cost structure supporting the existing communication equilibrium and the manner in which artificial intelligence alters these costs. Consider the case depicted in Figure~\ref{fig:equal-benefit}, where a signaling behavior distinguishes between types who incur high and low cognitive costs. If artificial intelligence enables the (formerly) high-cost types to effect the action at the same cost as the (formerly) low-cost types, then it will no longer differentiate types and its signaling power will collapse. However, if the introduction of artificial intelligence halves both costs, then some difference will remain and the behavior will retain some signaling power. Even in this latter case, the introduction of artificial intelligence will constrain the magnitude of benefits that can be made contingent upon the signal.

\section{Mechanism Two: Proofs of Knowledge}
 
Many forms of knowledge are difficult or even impossible to articulate \citep[so-called ``tacit'' knowledge;][]{polanyi1958personal,miton2022cultural}. In social contexts, this has the important implication that one cannot reveal certain relevant states of knowledge to others through mere recitation. This, in turn, can create barriers to cooperation, as many advantage social arrangements hinge on people possessing specific abilities (\emph{e.g.}, a contractual employment relationship requiring domain-specific expertise).
        
Fortunately, people can often prove they possess the underlying ability indirectly, by engaging in behaviors that would be prohibitively costly or impossible to perform without it. Beginning with \cite{goldwasser1985knowledge}, a sub-branch of cryptography has formally studied the properties of  protocols that enable one party (a ``prover'') to indirectly convince another party (a ``verifier'') that they possess a particular piece of knowledge (\emph{e.g.}, the solution to an intractable problem) or ability (\emph{e.g.}, can compute the values a specific function), without revealing much, if any, additional information beyond the mere possession of that knowledge or ability itself. 

Research into such protocols has clarified their essential structural features. According to framework originally laid out by \cite{goldwasser1985knowledge}, proof of knowledge protocols must satisfy two conditions: ``completeness'' (a sincere prover can always convince the verifier) and ``validity'' (a disingenuous prover cannot convince the verifier, except with arbitrarily small probability). 
        
Complete and valid protocols rely on the fact that knowledge enables us to do new things. Some actions would be improbable for someone who lacked the knowledge in question, and are therefore diagnostic of possessing it. For example, solving certain types of computationally intractable problems immediately entails solutions to a variety of other, structurally-related problems. Demonstrating capacity on these related problems convinces the verifier that the prover must, indeed, possess a solution to the original problem \citep{goldreich1991proofs,blum1986prove}. These insights have enabled a variety of applications, such as confidential voting \citep{groth2005non}, privacy-preserving machine learning \citep{minelli2018fully}, decentralized payments \citep{sasson2014zerocash}, and secure smart cards \citep{schnorr1990efficient}.
        
As \cite{goldwasser1985knowledge} point out, interactivity can afford distinct advantages: ``Writing, down a proof that can be checked by everybody without interaction is a much harder task. In some sense, because one has to answer in advance all possible questions'' (pg. 292).\footnotemark\ In the social domain, interactive proofs are, of course, familiar from standardized tests, where a short but probing evaluation of specific problems enables an examiner to verify a more general capacity. They are also essential to classroom instruction, where students ``may ask questions at crucial points of the argument and receive answers'' (\emph{Op. cit.}).

\footnotetext{Not all protocols, however, require interaction \cite{blum2019non}.}

Proof of knowledge exchanges extend well beyond the highly structured tests found in licensing exams and technical interviews, however: as we detail below, similar principles enable a sincere apology to certify that a wrongdoer does, indeed, possess a more elaborate mental model of their friend's needs and goals than the original \emph{faux pas} suggested. 

The validity of any such protocol, however, rests upon assumptions about the correlation between one's observable capabilities and possession of an underlying base of enabling knowledge. It is just this correlation that generative artificial intelligence disrupts: with access to a well-prompted machine, one can simulate many behaviors that would have previously required a far wider spectrum of personal abilities than just crafting a good prompt.

\section{Examples}
\subsection{A Simple Example: Sincere Apology}
\label{sec:apology}

In practice, many acts of mental proof in the social domain combine features of both costly signaling and proof of knowledge protocols to certify multiple mental facts simultaneously. 

Consider the ``sincere apology,'' a vital repair mechanism for human social relationships \citep{bachman2006forgiveness}. Although every transgressor would benefit from being forgiven, victims often have good reason not to: inductive evidence that the transgressor may behave poorly in the future. The very fact that apology follows a transgression means that contrary evidence must be communicated at a moment of low trust, hence the technology of mental proof is often vital to establishing one's credibility. 

To see why, consider the four points that \cite{lazare2005apology} lists as necessary for an effective apology: (1) identify both offending and offended parties; (2) acknowledge the incident in detail; (3) recognize the harms done; and (4) affirm that the behavior in question violated a social norm. To satisfy these points, the person apologizing must have a good mental model of the events in question and how they affected the injured party: in this sense, it is part of a proof of knowledge exchange. An explicit recitation of events not only establishes that the offending party knows what went wrong, but also generates common knowledge \citep{chwe2013rational} between both parties about the record of events and their significance relative to the injured party's needs and goals. 

Thinking through a situation in sufficient depth to have explored its various causes and consequences typically requires a significant, concerted expenditure of mental resources on cognitive operations such as mental simulation and perspective taking. One can demonstrate these efforts by making an apology particularly detailed and extensive, pairing it with a ``symbolic gesture,'' or by presenting novel insight into the situation. No matter the specific method employed, an effortful apology also serves as a costly signal that the offending party values the relationship enough to have invested in the process of rethinking and atoning for their actions.

A successful combination of proof of knowledge and costly signaling can provide strong evidence against the inductive hypothesis of continued transgression by proving that the offending party both: (1) understands the injured party's needs (proof of knowledge); and (2) values the relationship enough not to repeat the violation (costly signaling). Both parts are required: a heartfelt apology that misunderstands the transgression fails, no matter the effort, while an apology that appears easy and off-the-cuff fails, no matter how accurately it describes the situation.

Recognizing apology as an act of mental proof helps explain why a ``ChatGPT apology''\footnote{\url{https://abcnews.go.com/Business/chatgpt-wedding-vows-eulogies-stokes-dispute-authenticity/story?id=104763607}} written by artificial intelligence does not, to many, count as an apology at all. Recent experiments bear this out: \cite{glikson2023ai} find that people rate an apology as less sincere and are less likely to forgive when it was written using the heavy use of artificial intelligence tools. Interestingly, \citeauthor{glikson2023ai} also found that the light use of tools (\emph{e.g.}, to correct spelling or grammar errors) incurred no authenticity penalty, presumably because they did not meaningfully reduce perceived cognitive investment in the apology and, therefore, its capacity to deliver mental proof.




\subsection{A Complex Example: Social Proof}
\label{sec:social-facts}

Signaling and proof of knowledge reveal information about the preferences and proficiencies of the agents who engage them. By their very nature, therefore, mental proofs can only directly certify \emph{personal} facts about one's own mind. 

In certain contexts, however, individual facts combine to establish social facts. If a representative sample of people individually prove a mental fact, then an observer can statistically infer that it holds in a broader population. This is especially true in situations where coordination is unlikely or impossible.

Consider, for example, the powerful impact that encountering someone who can speak passionately about a highly specific interest that you, yourself happen to share. Such an encounter proves not just that you are not alone, but that there are many people like you. Groups establish not just identities, but common knowledge of their size and scope through mental proofs that involve hard-to-acquire knowledge of niche social facts, styles of speech, and shared beliefs.

The capacity of mental proofs to certify both personal and social facts  provides one explanation for the emergence of the baroque jargon and world-views in counter-normative communities, such as those devoted to conspiracy theories and extremist ideologies. While some jargon can serve as a shibboleth---a hard-to-forge marker of identity---other jargon is esoteric: learnable, certainly, but only with significant effort by true aficionados \citep{perry2021cognitive}. 

The role of mental proof in establishing social facts provides insight into recent work on artificially-enabled ``disinformation'' online. Naive observers, for example, who encounter increasing numbers of conspiracy-minded opinions online will naturally take that as evidence for the more general validity of the underlying belief. Such an effect, however, works only as long as the observers remain naive: once it becomes general knowledge that the behavior can be simulated with zero cost, we expect the opposite effect: the discounting of sincerely-professed beliefs and the reach of social movements that advance them.

\section{When Mental Proof Matters Most}

\begin{table*}[]
    \centering
    \resizebox{\textwidth}{!}{
    \begin{tabular}{p{1.8cm}|p{4.8cm}|p{4.8cm}|p{4.8cm}}

      \textbf{Activity} & \textbf{Mental fact demonstrated} & \textbf{AI Harm to Costly \mbox{Signaling}} & \textbf{AI Harm to Proof of \mbox{Knowledge}}  \\ \hline
      Apology (Section~\ref{sec:apology})  & ``They have contemplated how their actions affected me and now understand why they shouldn't behave similarly in the future.'' & ``They may have used an LLM to write this. If so, they do not actually care enough about me to think through the negative effects their actions had on me.'' & ``They may not actually understand why the actions harmed me or possess specific background knowledge necessary to avoid harming me in the future.'' \\ \hline
      Subcultural Membership (Section~\ref{sec:social-facts}) & ``Making an obscure reference shows that they have spent a lot of time with the artifacts of my subculture and are highly devoted to its associated values.'' & ``They may be using an LLM to imitate the  communication style of our community. They may have no investment in my subculture and are unlikely to share my values.'' & ``I cannot be sure if they know other facts or references associated with my subculture.'' \\ \hline
      Cold E-mail & ``They have demonstrated specific interest in working for my company through extensive internet research and can write intelligently about our industry.'' & ``They may have used an LLM to write similar messages to a wide array of companies and have no particular interest in my company or the problems it handles.'' & ``They may have used an LLM to simulate their expertise and do not possess the relevant underlying skills.'' \\ \hline
      Crisis Hotline \mbox{Volunteer} & ``Someone values me enough to have listened to my troubles and responded with encouragement.'' & ``They may have just fed my message into an LLM, which will return a generic responses regardless of what one says. I am truly alone.'' & ----- \\ \hline
      Technical Interview & ``By demonstrating obscure knowledge about an aspect of our domain, they show they are familiar enough with it that we will benefit from hiring them.'' & ----- & ``They may have gotten this fact from an LLM, without actually possessing any relevant background knowledge. They are not necessarily a good hire.''
    \end{tabular}
    }
    \caption{Examples of how generative AI can undermine the two principle mechanisms of mental proof: costly signaling and proof of knowledge.}
    \label{tab:examples_proof}
\end{table*}

Mental proofs are a pervasive features of social life; Table~\ref{tab:examples_proof} lists a few examples and describes how Generative AI undermines their social impact. Mental proofs are especially valuable in situations where both (1) deception is potentially profitable and (2) people cannot establish credibility in other ways. They are, therefore, particularly important before trust has been established at the beginning of a romantic, personal, or professional relationship.

Mental proofs also play an important role when when interactions are anonymous or infrequent, as happens in much online communication, or of such high stakes that a counter-party may be willing to sacrifice their reputation for an advantageous outcome. Finally, mental proofs are especially useful for communicating claims that are difficult to verify or articulate (\emph{e.g.}, abstract subjective claims such as caring or understanding).

Mental proofs are powerful, decentralized tools for building cooperation, but they are also, by their nature, costly; for this reason, societies often build more efficient mechanisms around, or on top of, them. If a society makes it possible for trading partners to interact over long periods of time, the accumulation of evidence can make costly proofs less necessary. Trusted reputation systems can ease interactions between strangers by providing a record of past acts of mental proof.

Mental proofs are most important in contexts where these supervening mechanisms are not present, for example because institutions are, or have become, weak or misaligned. Students for countries with failed credentialing systems, for example, will be asked to provide more mental proofs compared to others. Potential romantic partners may demand more mental proofs from each other in cultures that can not, or do not, enforce standards of care.

\section{When AI Damages Mental Proof}

Mental proofs are a powerful tool for establishing both individual relationships and the common knowledge necessary for effective group action. Conversely, the weakening of mental proof can significantly stunt people's ability to form new interpersonal relationships and cooperative initiatives. This suggests that the vibrancy of mental proof is a precursor to the general health of social, familial, political, and economic life in a community.

As the example of sincere apology makes clear, mental proof also play an important role in repairing and maintaining close ties. The general erosion of mental proof therefore also has the effect of diminishing the psychological and social benefits such relationships confer. Feeling that one is understood and cared about by another is a deep psychological priority \citep{cahn1990perceived,oishi2010felt,reis2000daily,lun2008feeling}, separate from the material benefits that care might bring.

Our discussion of social proof also showed how mental proofs help people establish collective interests, beliefs, and capacities. These proofs provide more than just the psychological benefits of knowing one is not alone. Collective action matters---the creation and maintenance of groups is a basic feature of civil society, in general, and the success of democratic government, in particular \citep{putnam1994making}.

\section{Implications}

Artificial intelligence's deleterious effects on both costly signaling and proofs of knowledge can be prevented if people can clearly delineate between communicative acts undertaken with and without the assistance of such tools. In the United States, this is a stated goal of a 2023 Executive Order (No. 14110, Section 4.5),\footnotemark\ and \cite{jain2023contextual} detail a variety of strategies to establish and maintain such distinctions, which they label the \emph{contextual confidence} of communication. 
Clarifying the difference between AI and human content would enable us to reap the benefits of automation (where advantageous) while preserving the capacity of humans to harness the benefits of mental proof (where important).

\footnotetext{\url{https://www.whitehouse.gov/briefing-room/presidential-actions/2023/10/30/executive-order-on-the-safe-secure-and-trustworthy-development-and-use-of-artificial-intelligence/}}

Our analysis also suggests that the economic benefits, in terms of reduced labor costs, of automating human mental effort can backfire. This may be especially true for jobs where care and understanding are paramount. Consider a patient in talk therapy for a difficult-to-treat condition. Generative artificial intelligence may help the therapist by spotting patterns and dynamically recommending better strategies of engagement, but we do not yet understand how this might undermine the therapeutic alliance \citep{alliance,alliance1}. Clarifying the work being done by mental proof may help guide efforts to surgically target aspects of these jobs that can benefit from automation without damaging their core efficacy.

Our work has focused on the beneficial effects of mental proof. As \cite{spence1973job} pointed out when introducing the concept, however, the existence of signaling equilibria can be socially costly and may create disadvantages for some participants. This suggests, in turn, that in some cases---which ones remains a subject for further research---outcomes may be improved when Generative AI destroys an equilibrium previously supported by mental proof. The conditions under which this is a net gain to the participants, either individually, or collectively, depends sensitively on the mechanisms that emerge to take its place.

Mental proof is most prominent in low-trust environments; consequently, its degradation stands to disproportionately impact those who already lack strong institutional support. Consider, for example, a student in the developing world without access to a functional accreditation system. In the past, a well-crafted, thoughtful e-mail might well serve to open doors to informal networks of mentorship and training: such a gesture provided both proof of abilities and of the necessary internal motivation that would lead a busy, but sympathetic, professor to take note. Such avenues to advancement are closed, however, once equivalent texts can be produced with a minimal prompt. A student without social capital is hurt by this vitiation of mental proof, while another, who comes with institutional backing, has other avenues to establish their credibility.

This suggests that expanding access to institutions that establish and sustain trust will be even more valuable for maintaining open and equatable societies as artificial intelligence advances. Novel protocols that bootstrap existing sources of trust \citep[\emph{e.g.},][]{weyl2022decentralized} could help counteract the inevitable changes that artificial intelligence will bring to the cost of thought and, through it, the structure of mental proof.

\bibliography{references}

\begin{thebibliography}{63}
\providecommand{\natexlab}[1]{#1}
\providecommand{\url}[1]{\texttt{#1}}
\expandafter\ifx\csname urlstyle\endcsname\relax
  \providecommand{\doi}[1]{doi: #1}\else
  \providecommand{\doi}{doi: \begingroup \urlstyle{rm}\Url}\fi

\bibitem[Glikson and Asscher(2023)]{glikson2023ai}
Ella Glikson and Omri Asscher.
\newblock {AI}-mediated apology in a multilingual work context: Implications for perceived authenticity and willingness to forgive.
\newblock \emph{Computers in Human Behavior}, 140:\penalty0 107592, 2023.

\bibitem[Fitria(2023)]{fitria2023artificial}
Tira~Nur Fitria.
\newblock {Artificial intelligence (AI) technology in OpenAI ChatGPT application: A review of ChatGPT in writing English essay}.
\newblock In \emph{ELT Forum: Journal of English Language Teaching}, volume~12, pages 44--58, 2023.

\bibitem[Cardon et~al.(2023)Cardon, Fleischmann, Aritz, Logemann, and Heidewald]{cardon2023challenges}
Peter Cardon, Carolin Fleischmann, Jolanta Aritz, Minna Logemann, and Jeanette Heidewald.
\newblock The challenges and opportunities of ai-assisted writing: Developing ai literacy for the ai age.
\newblock \emph{Business and Professional Communication Quarterly}, 86\penalty0 (3):\penalty0 257--295, 2023.

\bibitem[Wu and Kelly(2020)]{wu2020online}
Yihan Wu and Ryan~M Kelly.
\newblock Online dating meets artificial intelligence: How the perception of algorithmically generated profile text impacts attractiveness and trust.
\newblock In \emph{Proceedings of the 32nd Australian Conference on Human-Computer Interaction}, pages 444--453, 2020.

\bibitem[LaGorce(2023)]{nyt}
Tammy LaGorce.
\newblock Need to write your vows? {A.I.} can help.
\newblock 2023.
\newblock URL \url{https://www.nytimes.com/2023/03/03/fashion/weddings/chatbot-wedding-vows-chatgpt-ai.html}.

\bibitem[Glikson and Woolley(2020)]{glikson2020human}
Ella Glikson and Anita~Williams Woolley.
\newblock Human trust in artificial intelligence: Review of empirical research.
\newblock \emph{Academy of Management Annals}, 14\penalty0 (2):\penalty0 627--660, 2020.

\bibitem[Jain et~al.(2023)Jain, Hitzig, and Mishkin]{jain2023contextual}
Shrey Jain, Zo{\"e} Hitzig, and Pamela Mishkin.
\newblock Contextual confidence and generative ai.
\newblock \emph{arXiv preprint arXiv:2311.01193}, 2023.

\bibitem[Leslie(2019)]{leslie2019understanding}
David Leslie.
\newblock Understanding artificial intelligence ethics and safety.
\newblock \emph{arXiv preprint arXiv:1906.05684}, 2019.

\bibitem[Allen and Weyl(2024)]{allen2024real}
Danielle Allen and E~Glen Weyl.
\newblock The real dangers of generative ai.
\newblock \emph{Journal of Democracy}, 35\penalty0 (1):\penalty0 147--162, 2024.

\bibitem[Jungherr(2023)]{jungherr2023artificial}
Andreas Jungherr.
\newblock Artificial intelligence and democracy: A conceptual framework.
\newblock \emph{Social media+ society}, 9\penalty0 (3):\penalty0 20563051231186353, 2023.

\bibitem[Tyson(2023)]{Tyson_2023}
Alec Tyson.
\newblock Growing public concern about the role of artificial intelligence in daily life, Aug 2023.
\newblock URL \url{https://www.pewresearch.org/short-reads/2023/08/28/growing-public-concern-about-the-role-of-artificial-intelligence-in-daily-life/}.

\bibitem[Solaiman et~al.(2023)Solaiman, Talat, Agnew, Ahmad, Baker, Blodgett, Daum{\'e}~III, Dodge, Evans, Hooker, et~al.]{solaiman2023evaluating}
Irene Solaiman, Zeerak Talat, William Agnew, Lama Ahmad, Dylan Baker, Su~Lin Blodgett, Hal Daum{\'e}~III, Jesse Dodge, Ellie Evans, Sara Hooker, et~al.
\newblock Evaluating the social impact of generative ai systems in systems and society.
\newblock \emph{arXiv preprint arXiv:2306.05949}, 2023.

\bibitem[Weidinger et~al.(2021)Weidinger, Mellor, Rauh, Griffin, Uesato, Huang, Cheng, Glaese, Balle, Kasirzadeh, et~al.]{weidinger2021ethical}
Laura Weidinger, John Mellor, Maribeth Rauh, Conor Griffin, Jonathan Uesato, Po-Sen Huang, Myra Cheng, Mia Glaese, Borja Balle, Atoosa Kasirzadeh, et~al.
\newblock Ethical and social risks of harm from language models.
\newblock \emph{arXiv preprint arXiv:2112.04359}, 2021.

\bibitem[Mirsky and Lee(2021)]{mirsky2021creation}
Yisroel Mirsky and Wenke Lee.
\newblock The creation and detection of deepfakes: A survey.
\newblock \emph{ACM Computing Surveys (CSUR)}, 54\penalty0 (1):\penalty0 1--41, 2021.

\bibitem[Tomasello et~al.(2005)Tomasello, Carpenter, Call, Behne, and Moll]{tomasello2005understanding}
Michael Tomasello, Malinda Carpenter, Josep Call, Tanya Behne, and Henrike Moll.
\newblock Understanding and sharing intentions: The origins of cultural cognition.
\newblock \emph{Behavioral and brain sciences}, 28\penalty0 (5):\penalty0 675--691, 2005.

\bibitem[Gilbert(1990)]{gilbert1990walking}
Margaret Gilbert.
\newblock Walking together: A paradigmatic social phenomenon.
\newblock \emph{MidWest studies in philosophy}, 15\penalty0 (1):\penalty0 1--14, 1990.

\bibitem[Bratman(1992)]{bratman1992shared}
Michael~E Bratman.
\newblock Shared cooperative activity.
\newblock \emph{The philosophical review}, 101\penalty0 (2):\penalty0 327--341, 1992.

\bibitem[Farrell(1987)]{farrell1987cheap}
Joseph Farrell.
\newblock Cheap talk, coordination, and entry.
\newblock \emph{The RAND Journal of Economics}, pages 34--39, 1987.

\bibitem[Fehr(2009)]{fehr2009economics}
Ernst Fehr.
\newblock On the economics and biology of trust.
\newblock \emph{Journal of the european economic association}, 7\penalty0 (2-3):\penalty0 235--266, 2009.

\bibitem[Elster(1989)]{elster1989social}
Jon Elster.
\newblock Social norms and economic theory.
\newblock \emph{Journal of economic perspectives}, 3\penalty0 (4):\penalty0 99--117, 1989.

\bibitem[North(1991)]{north1991institutions}
Douglass~C North.
\newblock Institutions.
\newblock \emph{Journal of economic perspectives}, 5\penalty0 (1):\penalty0 97--112, 1991.

\bibitem[Spence(1973)]{spence1973job}
Michael Spence.
\newblock Job market signaling.
\newblock \emph{The Quarterly Journal of Economics}, 87\penalty0 (3):\penalty0 355--374, 1973.

\bibitem[Zahavi(1975)]{zahavi1975mate}
Amotz Zahavi.
\newblock Mate selection—a selection for a handicap.
\newblock \emph{Journal of theoretical Biology}, 53\penalty0 (1):\penalty0 205--214, 1975.

\bibitem[Smith and Harper(2003)]{smith2003animal}
John~Maynard Smith and David Harper.
\newblock \emph{Animal signals}.
\newblock Oxford University Press, 2003.

\bibitem[Fitzgibbon and Fanshawe(1988)]{fitzgibbon1988stotting}
Claire~D Fitzgibbon and John~H Fanshawe.
\newblock Stotting in thomson's gazelles: an honest signal of condition.
\newblock \emph{Behavioral Ecology and Sociobiology}, 23:\penalty0 69--74, 1988.

\bibitem[Caro(1994)]{caro1994ungulate}
Tim~M Caro.
\newblock Ungulate antipredator behaviour: preliminary and comparative data from african bovids.
\newblock \emph{Behaviour}, 128\penalty0 (3-4):\penalty0 189--228, 1994.

\bibitem[Milgrom and Roberts(1986)]{milgrom1986price}
Paul Milgrom and John Roberts.
\newblock Price and advertising signals of product quality.
\newblock \emph{Journal of political economy}, 94\penalty0 (4):\penalty0 796--821, 1986.

\bibitem[Cheng et~al.(2014)Cheng, Ioannou, and Serafeim]{cheng2014corporate}
Beiting Cheng, Ioannis Ioannou, and George Serafeim.
\newblock Corporate social responsibility and access to finance.
\newblock \emph{Strategic management journal}, 35\penalty0 (1):\penalty0 1--23, 2014.

\bibitem[Flammer(2021)]{flammer2021corporate}
Caroline Flammer.
\newblock Corporate green bonds.
\newblock \emph{Journal of financial economics}, 142\penalty0 (2):\penalty0 499--516, 2021.

\bibitem[Ahlers et~al.(2015)Ahlers, Cumming, G{\"u}nther, and Schweizer]{ahlers2015signaling}
Gerrit~KC Ahlers, Douglas Cumming, Christina G{\"u}nther, and Denis Schweizer.
\newblock Signaling in equity crowdfunding.
\newblock \emph{Entrepreneurship theory and practice}, 39\penalty0 (4):\penalty0 955--980, 2015.

\bibitem[Henrich et~al.(2010)Henrich, Heine, and Norenzayan]{weird}
Joseph Henrich, Steven~J Heine, and Ara Norenzayan.
\newblock Most people are not {WEIRD}.
\newblock \emph{Nature}, 466\penalty0 (7302):\penalty0 29--29, 2010.

\bibitem[Bird and Smith(2005)]{bliegebird2005signaling}
Rebecca Bird and Eric Smith.
\newblock Signaling theory, strategic interaction, and symbolic capital.
\newblock \emph{Current anthropology}, 46\penalty0 (2):\penalty0 221--248, 2005.

\bibitem[Bergstrom and Lachmann(2001)]{bergstrom2001alarm}
Carl~T Bergstrom and Michael Lachmann.
\newblock Alarm calls as costly signals of antipredator vigilance: the watchful babbler game.
\newblock \emph{Animal behaviour}, 61\penalty0 (3):\penalty0 535--543, 2001.

\bibitem[Mithen(2006)]{mithen2006singing}
Steven~J Mithen.
\newblock \emph{The singing Neanderthals: The origins of music, language, mind, and body}.
\newblock Harvard University Press, 2006.

\bibitem[Folstad and Karter(1992)]{folstad1992parasites}
Ivar Folstad and Andrew~John Karter.
\newblock Parasites, bright males, and the immunocompetence handicap.
\newblock \emph{The American Naturalist}, 139\penalty0 (3):\penalty0 603--622, 1992.

\bibitem[Miller(1956)]{miller1956magical}
George~A Miller.
\newblock The magical number seven, plus or minus two: Some limits on our capacity for processing information.
\newblock \emph{Psychological review}, 63\penalty0 (2):\penalty0 81, 1956.

\bibitem[Shenhav et~al.(2017)Shenhav, Musslick, Lieder, Kool, Griffiths, Cohen, and Botvinick]{shenhav2017toward}
Amitai Shenhav, Sebastian Musslick, Falk Lieder, Wouter Kool, Thomas~L Griffiths, Jonathan~D Cohen, and Matthew~M Botvinick.
\newblock Toward a rational and mechanistic account of mental effort.
\newblock \emph{Annual review of neuroscience}, 40:\penalty0 99--124, 2017.

\bibitem[Loewenstein and Wojtowicz(2023)]{loewenstein2023economics}
George Loewenstein and Zachary Wojtowicz.
\newblock The economics of attention.
\newblock \emph{Available at SSRN 4368304}, 2023.

\bibitem[Nordhaus(1996)]{nordhaus1996real}
William~D Nordhaus.
\newblock Do real-output and real-wage measures capture reality? the history of lighting suggests not.
\newblock In \emph{The economics of new goods}, pages 27--70. University of Chicago Press, 1996.

\bibitem[Simon(1971)]{simon1971designing}
Herbert~A. Simon.
\newblock Designing organizations for an information-rich world.
\newblock 1971.

\bibitem[Attwell and Laughlin(2001)]{attwell2001energy}
David Attwell and Simon~B Laughlin.
\newblock An energy budget for signaling in the grey matter of the brain.
\newblock \emph{Journal of Cerebral Blood Flow \& Metabolism}, 21\penalty0 (10):\penalty0 1133--1145, 2001.

\bibitem[Polanyi(1958)]{polanyi1958personal}
Michael Polanyi.
\newblock Personal knowledge: Towards a post-critical philosophy.
\newblock 1958.

\bibitem[Miton and DeDeo(2022)]{miton2022cultural}
Helena Miton and Simon DeDeo.
\newblock The cultural transmission of tacit knowledge.
\newblock \emph{Journal of the Royal Society Interface}, 19\penalty0 (195):\penalty0 20220238, 2022.

\bibitem[Goldwasser et~al.(1985)Goldwasser, Micali, and Rackoff]{goldwasser1985knowledge}
S~Goldwasser, S~Micali, and C~Rackoff.
\newblock The knowledge complexity of interactive proof-systems.
\newblock In \emph{Proceedings of the seventeenth annual ACM symposium on Theory of computing}, pages 291--304, 1985.

\bibitem[Goldreich et~al.(1991)Goldreich, Micali, and Wigderson]{goldreich1991proofs}
Oded Goldreich, Silvio Micali, and Avi Wigderson.
\newblock Proofs that yield nothing but their validity or all languages in np have zero-knowledge proof systems.
\newblock \emph{Journal of the ACM (JACM)}, 38\penalty0 (3):\penalty0 690--728, 1991.

\bibitem[Blum(1986)]{blum1986prove}
Manuel Blum.
\newblock How to prove a theorem so no one else can claim it.
\newblock In \emph{Proceedings of the International Congress of Mathematicians}, volume~1, page~2. Citeseer, 1986.

\bibitem[Groth(2005)]{groth2005non}
Jens Groth.
\newblock Non-interactive zero-knowledge arguments for voting.
\newblock In \emph{Applied Cryptography and Network Security: Third International Conference, ACNS 2005, New York, NY, USA, June 7-10, 2005. Proceedings 3}, pages 467--482. Springer, 2005.

\bibitem[Minelli(2018)]{minelli2018fully}
Michele Minelli.
\newblock \emph{Fully homomorphic encryption for machine learning}.
\newblock PhD thesis, Universit{\'e} Paris sciences et lettres, 2018.

\bibitem[Sasson et~al.(2014)Sasson, Chiesa, Garman, Green, Miers, Tromer, and Virza]{sasson2014zerocash}
Eli~Ben Sasson, Alessandro Chiesa, Christina Garman, Matthew Green, Ian Miers, Eran Tromer, and Madars Virza.
\newblock Zerocash: Decentralized anonymous payments from bitcoin.
\newblock In \emph{2014 IEEE symposium on security and privacy}, pages 459--474. IEEE, 2014.

\bibitem[Schnorr(1990)]{schnorr1990efficient}
Claus-Peter Schnorr.
\newblock Efficient identification and signatures for smart cards.
\newblock In \emph{Advances in Cryptology—CRYPTO’89 Proceedings 9}, pages 239--252. Springer, 1990.

\bibitem[Blum et~al.(2019)Blum, Feldman, and Micali]{blum2019non}
Manuel Blum, Paul Feldman, and Silvio Micali.
\newblock Non-interactive zero-knowledge and its applications.
\newblock In \emph{Providing Sound Foundations for Cryptography: On the Work of Shafi Goldwasser and Silvio Micali}, pages 329--349. 2019.

\bibitem[Bachman and Guerrero(2006)]{bachman2006forgiveness}
Guy~Foster Bachman and Laura~K Guerrero.
\newblock Forgiveness, apology, and communicative responses to hurtful events.
\newblock \emph{Communication reports}, 19\penalty0 (1):\penalty0 45--56, 2006.

\bibitem[Lazare(2005)]{lazare2005apology}
Aaron Lazare.
\newblock \emph{On apology}.
\newblock Oxford University Press, 2005.

\bibitem[Chwe(2013)]{chwe2013rational}
Michael Suk-Young Chwe.
\newblock \emph{Rational ritual: Culture, coordination, and common knowledge}.
\newblock Princeton University Press, 2013.

\bibitem[Perry and DeDeo(2021)]{perry2021cognitive}
Chloe Perry and Simon DeDeo.
\newblock The cognitive science of extremist ideologies online.
\newblock \emph{arXiv}, 2110.00626, 2021.

\bibitem[Cahn(1990)]{cahn1990perceived}
Dudley~D Cahn.
\newblock Perceived understanding and interpersonal relationships.
\newblock \emph{Journal of Social and Personal Relationships}, 7\penalty0 (2):\penalty0 231--244, 1990.

\bibitem[Oishi et~al.(2010)Oishi, Krochik, and Akimoto]{oishi2010felt}
Shigehiro Oishi, Margarita Krochik, and Sharon Akimoto.
\newblock Felt understanding as a bridge between close relationships and subjective well-being: Antecedents and consequences across individuals and cultures.
\newblock \emph{Social and Personality Psychology Compass}, 4\penalty0 (6):\penalty0 403--416, 2010.

\bibitem[Reis et~al.(2000)Reis, Sheldon, Gable, Roscoe, and Ryan]{reis2000daily}
Harry~T Reis, Kennon~M Sheldon, Shelly~L Gable, Joseph Roscoe, and Richard~M Ryan.
\newblock Daily well-being: The role of autonomy, competence, and relatedness.
\newblock \emph{Personality and social psychology bulletin}, 26\penalty0 (4):\penalty0 419--435, 2000.

\bibitem[Lun et~al.(2008)Lun, Kesebir, and Oishi]{lun2008feeling}
Janetta Lun, Selin Kesebir, and Shigehiro Oishi.
\newblock On feeling understood and feeling well: The role of interdependence.
\newblock \emph{Journal of Research in Personality}, 42\penalty0 (6):\penalty0 1623--1628, 2008.

\bibitem[Putnam(1994)]{putnam1994making}
Robert~D Putnam.
\newblock \emph{Making democracy work: Civic traditions in modern Italy}.
\newblock Princeton University Press, 1994.

\bibitem[Tal et~al.(2023)Tal, Elyoseph, Haber, Angert, Gur, Simon, and Asman]{alliance}
Amir Tal, Zohar Elyoseph, Yuval Haber, Tal Angert, Tamar Gur, Tomer Simon, and Oren Asman.
\newblock The artificial third: utilizing chatgpt in mental health.
\newblock \emph{The American Journal of Bioethics}, 23\penalty0 (10):\penalty0 74--77, 2023.

\bibitem[Zetzel(1956)]{alliance1}
Elizabeth~R. Zetzel.
\newblock An approach to the relation between concept and content in psychoanalytic theory.
\newblock \emph{The Psychoanalytic Study of the Child}, 11\penalty0 (1):\penalty0 99--121, 1956.
\newblock \doi{10.1080/00797308.1956.11822784}.

\bibitem[Weyl et~al.(2022)Weyl, Ohlhaver, and Buterin]{weyl2022decentralized}
E~Glen Weyl, Puja Ohlhaver, and Vitalik Buterin.
\newblock Decentralized society: Finding web3's soul.
\newblock \emph{Available at SSRN 4105763}, 2022.

\end{thebibliography}

\end{document}